\documentclass[10pt, conference, letterpaper]{IEEEtran}
\usepackage{balance}
\usepackage{graphicx}
\usepackage{subfigure}

\usepackage{booktabs,multirow}

\usepackage{algorithm}
\usepackage{algpseudocode}
\usepackage{color}

\usepackage{xspace}
\usepackage{url}

\usepackage[inline]{enumitem}

\newcommand{\secref}[1]{{\S\ref{#1}}}
\newcommand{\fref}[1]{\mbox{Figure~\ref{#1}}}

\newcommand{\aref}[1]{\mbox{Algorithm~\ref{#1}}}

\newcommand{\myparab}[1]{\vspace{0.075in}\noindent\textbf{#1}\xspace}

\newcommand{\eg}{{\it e.g.}\xspace}
\newcommand{\ie}{{\it i.e.}\xspace}

\newcommand{\eat}[1]{}

\newcommand{\algo}{\textsc{Scout}\xspace}
\newcommand{\score}{{SCORE}\xspace}
\newcommand{\system}{\textsc{Scout}\xspace}

\usepackage{mathptmx}
\usepackage[mathcal]{euscript}
\usepackage[T1]{fontenc}

\newcommand{\subparagraph}{}

\newlength{\textfloatsepsave}
\begin{document}

\date{}

\title{Fault Localization in Large-Scale Network Policy Deployment}

 \author{Praveen Tammana$^\dag$  \quad Chandra Nagarajan$^\ddag$ \quad Pavan Mamillapalli$^\ddag$ \quad  Ramana Rao Kompella$^\ddag$ \\ \quad Myungjin
  Lee$^\dag$  \\
 \small {\em $^\dag$University of Edinburgh, $^\ddag$Cisco Systems}
 }

\maketitle

\begin{abstract}
The recent advances in network management automation and Software-Defined
Networking (SDN) are easing network policy management tasks. At the same time,
these new technologies create a new mode of failure in the management cycle
itself. Network policies are presented in an abstract model at a centralized
controller and deployed as low-level rules across network devices. Thus, any
software and hardware element in that cycle can be a potential cause of
underlying network problems. In this paper, we present and solve a \emph{network
policy fault localization problem} that arises in operating policy management
frameworks for a production network.  We formulate our problem via risk modeling
and propose a greedy algorithm that quickly localizes faulty policy objects in
the network policy. We then design and develop \system---a fully-automated
system that produces faulty policy objects and further pinpoints physical-level
failures which made the objects faulty.  Evaluation results using a real testbed
and extensive simulations demonstrate that \system detects faulty objects with
small false positives and false negatives.
\end{abstract}

\section{Introduction}
\label{s:intro}

Fast fault localization in the network is essential but becoming more
challenging than ever. Modern networks are increasingly complex. The network
infrastructures support new complex functions such as virtualization,
multitenancy, performance isolation, access control, and so forth.
The instantiation of these functions is governed by high-level network policies
that reflect on network-wide requirements. SDN\footnote{In this paper, SDN is
used in a broader context, not just limited to the OpenFlow protocol, to which
all the discussions are still relevant.} makes such network management tasks
easier with a global view on the network state at a logically centralized SDN
controller.

In a network, a vast amount of low-level configuration instructions can be
translated from a few high-level policies. Errors that lurk during policy
creation, translation or delivery, may lead to the incorrect deployment of a
large number of low-level rules in network devices. A single error for a policy
can cause a serious damage such as outage to business-critical services. Hence,
the network policy management process of SDN creates a new mode of failure.

A number of frameworks~\cite{pga, merlin, apic, frenetic, pyretic, maple, gbp}
aid network policy management tasks through abstraction, policy composition and
deployment. However, these frameworks are not immune to various faulty
situations that can arise from misconfiguration, software bugs, hardware
failure, control channel disruption, device memory overflow, etc. Many of them
incur a flow of instructions from a centralized controller, to a software agent
in a network device and finally to ternary content addressable memory (TCAM) in
that device. Thus, any element in this data flow can be the root cause of policy
deployment failures.

When a network policy is not rendered in the network as expected, network admins
should first understand which part of the policy has been affected. This is
challenging because the admins can end up examining tens of thousands of
low-level rules. In the existing policy management frameworks~\cite{pga, apic},
low-level rules are built from policy objects (in short, objects) such as
marketing group, DB tier, filter, and so on. Our study on a production cluster
reveals that even one object can be used to create TCAM rules for over thousands
of endpoints (\secref{s:cardinality} and \fref{fig:cardinality}). This implies
that a fault of that single object can lead to a communication outage for those
numerous endpoints, and the admins observe too many failures. 
Examining all of the TCAM rules associated with the endpoints would be tedious.
Thus, the admins require a fully-automated means that quickly nails down to the
part of the policy they should look into or further diagnose to fix a large
number of the observed failures.

We call this problem of finding out the impaired parts of the policy as a
\emph{network policy fault localization problem}, which we tackle via \emph{risk modeling}~\cite{score}. We model risks as simple
bipartite graphs that capture dependencies between risks (\ie, objects) and
nodes (\eg, endpoints or end user applications) that rely on those risks. We
then annotate the risk models for those risks and nodes associated with the
observed failures. Using the annotated risk models, we devise a greedy fault
localization algorithm that outputs a hypothesis, a minimum set of most-likely
faulty policy objects (\ie, risks) that explains most of the observed failures.

At first glance, solving this policy fault localization problem looks
straightforward as a similar problem has been studied for IP
networks~\cite{score}. However, as this paper tries to address the problem in a
new operating domain --- SDN-enabled data center and enterprise networks, there
are two key challenges. First, it is difficult to represent risks in the network
policy as a single model. Solving many risk models can be computationally
expensive. In our modeling, we fortunately require two risk models only: switch
risk model and controller risk model (\secref{sec:risk_models}). We make the two
models based on our observation that faults of policy objects occur at two broad
layers (controller and switch). If the controller malfunctions, unsuccessful
policy deployment can potentially affect all the switches in the network (thus,
controller risk model). In contrast, a policy deployment failure can be limited
to a switch if that switch only becomes faulty (thus, switch risk model).

\begin{figure*}[t]
\centering
\includegraphics[width=0.9\textwidth]{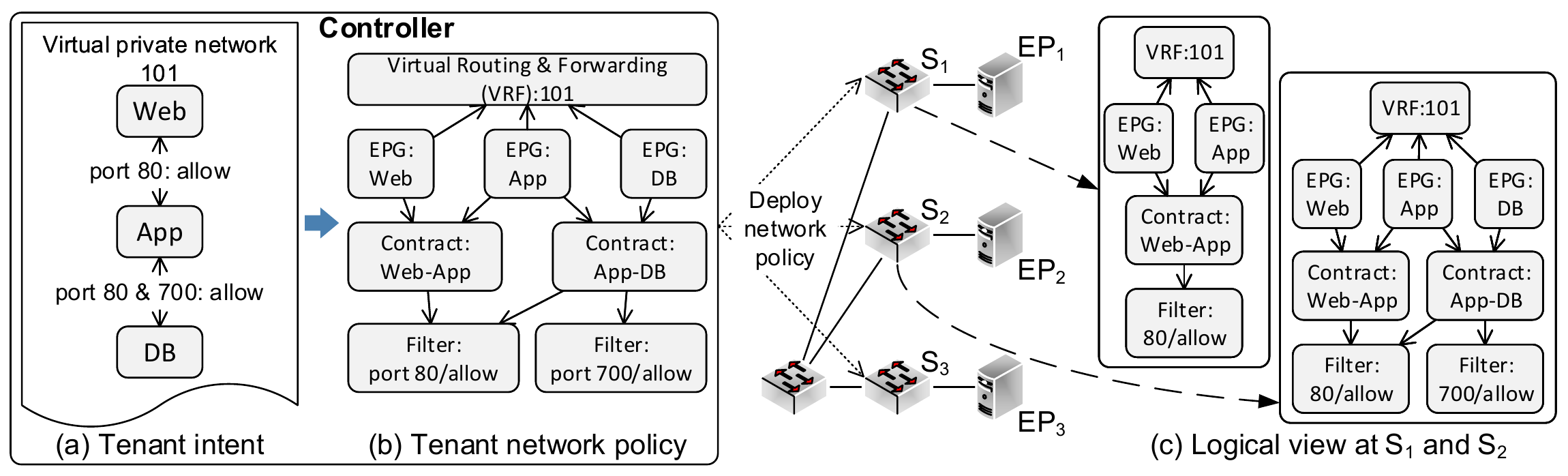}
\caption{An example of network policy management framework. EP stands for
  endpoint, and EPG denotes endpoint group.}
\label{fig:bgexp}
\end{figure*}

Our second challenge stems from the fact that the degree of impact on endpoints
caused by a faulty object varies substantially.  When a fault event occurs, 
a subset of objects are responsible for all of the impacted endpoints. In contrast, another subset of objects cause trouble to a small fraction of the whole endpoints that rely on them. This variety makes accurate fault localization difficult. An existing algorithm~\cite{score} tends to choose policy objects in the former case while
it treats objects in the latter case as input noise.
However, in our problem, some objects do belong to the latter case. To handle
this issue, \system employs a 2-stage approach;
it first picks objects only if all of their dependent endpoints are in failure;
next, for (typically a small number of) objects left unexplained in the risk
model, it looks up the change logs (maintained at a controller) and selects the
objects to which some actions are recently applied (\secref{sec:faultloc}).
Despite its simplicity, this heuristic effectively localizes faulty objects.
(\secref{sec:eval}).

Overall, this paper makes the following main contributions.
\begin{enumerate}
\setlength\itemsep{0.5em}
\item We introduce and study a network policy fault localization problem
  (\secref{sec:background}) in the context of SDN. This is a new problem that
  gained little attention but is of utmost importance in operating a network
  policy management framework safely.

\item We introduce two risk models (switch and controller risk models) that
  precisely capture the characteristics of the problem and help its formulation
  (\secref{s:risk}).
\item We devise a network policy fault localization algorithm that quickly
  narrows down a small number of suspicious faulty objects
  (\secref{sec:faultloc}). We then design and implement \system
  (\secref{sec:arch}), a system that conducts an end-to-end automatic fault
  localization from failures on policy objects to physical-level failures that
  made the objects faulty.

\item We evaluate \system using a real production cluster and extensive
  simulations (\secref{sec:eval}). Our evaluations show that \system achieves
  20-50\% higher accuracy than an existing solution and is scalable. \system
  runs a large-scale controller risk model of a network with 500 leaf switches,
  under 130~seconds in a commodity machine.
\end{enumerate}

\section{Policy deployment by example} 
\label{sec:background}

In this section we first introduce network policy, its abstraction model, and
its deployment. We then discuss network state inconsistency caused by failures
of elements involved in the network policy management.

\subsection{Network policy}

In general network policies dictate the way traffic should be treated in a
network. In managing network policies, tenant/admins should be able to express
their intent on traffic via a model and to enforce the policies at individual
network devices. To enable more flexible composition and management of network
policies, several frameworks~\cite{pga, apic, gbp} present the network policies
in an abstracted model (\eg, a graph) that describes communication relationships
among phyiscal/logical entities such as servers, switches, middleboxes, VMs,
etc.

\myparab{Intent illustration.} As an example, consider a canonical 3-tier web
service that consists of Web, App and DB servers (or VMs) as shown in
\fref{fig:bgexp}(a). Here the tenant intent is to allow communication on specific ports between the application tiers, \ie, port 80 between Web and App, ports 80 and 700 between App and DB. A network policy framework transforms intent of users (tenant,
network admins, etc.) into an abstracted policy as illustrated in
\fref{fig:bgexp}(b).

\myparab{Network policy presentation.} For driving our discussion, we here apply
a network policy abstraction model used in Cisco's application-centric policy
infrastructure controller (APIC)~\cite{apic}, which is quite similar to other
models (\eg, GBP~\cite{gbp}, PGA~\cite{pga}); and our work for localizing faults
in network policy management is agnostic to policy abstraction model itself.
\fref{fig:bgexp}(b) illustrates a network policy (as a graph represented with
policy objects) transformed from the tenant intent shown in \fref{fig:bgexp}(a).
We discuss each of those policy objects next.

An \emph{endpoint group (EPG)} represents a set of \emph{endpoints} (EPs), \eg,
servers, VMs, and middleboxes, that belong to the same application tier. A
\emph{filter} governs access control between EPGs. This policy entity takes a
whitelisting apporach, which by default blocks all traffic in the absence of
mapping between EPGs and filters.

The mapping between EPGs and filters are indirectly managed by an object called
\emph{contract}, which serves as a glue between EPGs and filters. A contract
defines what filters need to be applied to which EPGs. Thus, a contract enables
easy modification of filters. For example, in \fref{fig:bgexp}(b), let us assume
EPG:App and EPG:DB no longer require communication between them on port
700. This only requires to remove ``Filter: port 700/allow'' from the
Contract:App-DB; no update between EPGs and their contract is necessary.

Finally, the scope of all EPGs in a tenant policy is defined using a layer-3
virtual private network, realized with a virtual routing and forwarding
(VRF) object.

\myparab{Network policy deployment.} A network policy should be realized through
deployment. A centralized controller maintains the network policy and makes
changes on it. When updates (add/delete/modify) on a network policy are made,
the controller compiles the new policy and produces instructions that consist of
policy objects and the update operations associated with the objects. The
controller then distributes the instructions to respective switch agents. The
switch agents also keep a local view on the network policy to which the
instructions from the controller are applied. The switch agents transform any
changes on the logical view into low-level TCAM rules. Note that there are
multiple technologies to link controller and switch agents like OpenFlow,
OpFlex~\cite{opflex}, etc. Also, TCAM could have matching rules based on either
typical IP packet header fields or custom proprietary header fields. Our work
on fault localization is agnostic to both linking technologies and the format of
TCAM rules.

\begin{figure}[t]
\small
\centering
  \begin{tabular}{@{}crc@{}} \toprule
    \textbf{No.} & \multicolumn{1}{c}{\textbf{Rule}} & \textbf{Action} \\ \midrule
    1 & VRF:101,Web,App,Port80 & Allow \\
    2 & VRF:101,App,Web,Port80 & Allow \\
    3 & VRF:101,App,DB,Port80 & Allow \\
    4 & VRF:101,DB,App,Port80 & Allow \\
    5 & VRF:101,App,DB,Port700 & Allow \\
    6 & VRF:101,DB,App,Port700 & Allow \\
    7 & \multicolumn{1}{c}{{*},*,*,*} & Deny \\
\bottomrule
 \end{tabular} 
 
 \caption{TCAM rules in switch $S_2$. Note that here a rule is annotated with
    object types in it for ease of exposition.
  }
\label{fig:swrules}
\end{figure}

Consider a network topology (\fref{fig:bgexp}) where $EP_1$ is attached to
switch $S_1$, $EP_2$ to $S_2$ and $EP_3$ to $S_3$. Let us assume that
$EP_1 \in EPG\mathrm{:}Web$, $EP_2 \in EPG\mathrm{:}App$ and
$EP_3 \in EPG\mathrm{:}DB$. Putting it altogether, the controller sends out the
instructions about $EPG\mathrm{:}Web$ to switch $S_1$ (as $EP_1$ is connected to
$S_1$), those about $EPG\mathrm{:}App$ to switch $S_2$, and so forth. As the
three switches receive the instructions on those EPGs for the first time, they
build a logical view from scratch (see \fref{fig:bgexp}(c) for example). Hence,
a series of add operations invoke TCAM rule installations in each
switch. \fref{fig:swrules} shows access control list (ACL) rules rendered in
TCAM of $S_2$.

\subsection{Network state inconsistency}

Network policy enforcement is by nature a distributed process and involves the
management of three key elements: (i) a global network policy at controller,
(ii) a local network policy at switch agent, and (iii) TCAM rules generated from
the local policy. Ideally, the states among these three elements should be
\emph{equivalent} in order for the network to function as intended by admins.

In reality, these elements may not be in an equivalent state due to a number of
reasons. A switch agent may crash in the middle of TCAM rule updates.  A
temporal disconnection between the controller and switch agent during the
instruction push. TCAM has insufficient space to add new ACL rules, which
renders the rule installation incomplete. The agent may run a local rule
eviction mechanism, which even worsens the situation because the controller may
be unaware of the rules deleted from TCAM. Even TCAM is simply corrupted due to
hardware failure. All of these cases can create a state mismatch among
controller, switch agent and TCAM level, which compromises the integrity of the
network.

One approach to this issue is to make network policy management frameworks more
resilient against failures. However, failures are inevitable, so is the network
state inconsistency.

\section{Shared Risks in Network Policy}
\label{s:risk}

We exploit shared risk models for our network policy fault localization
problem. The shared risk model has been well studied in IP
networks~\cite{score}.
For instance, when a fiber optic cable carries multiple logical IP links, the
cable is recognized as a \emph{shared risk} for those IP links because the
optical cable failure would make the IP links fail or perform poorly.

Deploying a network policy also presents shared risks. A network policy
comprises policy objects (such as VRF, EPGs, contract, filter, etc). The
relationship among those objects dictates how a network policy must be realized.
If an object is absent or ill-represented in any of controller, switch agent and
TCAM layers, all of EPG pairs that rely on that object would be negatively
impacted. Thus, these policy objects on which a set of EPG pairs rely are
shared risks in the network policy deployment.

\fref{fig:swrules} depicts that a TCAM rule is expressed as a combination of
objects presented in a logical model at switch $S_2$. If the 5th and 6th TCAM
rules in the figure are absent from TCAM, all the traffic between EPG:App and
EPG:DB via port 700 would be dropped. The absence of correct rules boils down to
a case where one or more objects are not rendered correctly in TCAM; a corrupted
TCAM may write a wrong VRF identifier (ID) or EPG ID for those rules; $S_2$ may
drop the filter `port 700/allow' from its logical view due to software bug. Such
absence or mispresentation of objects directly affect the EPG pairs that
share the objects. Thus, shared risk objects for App-DB EPG pair are VRF:101,
EPG:App, EPG:DB, Contract:App-DB, Filter:80/allow, and Filter:700/allow.

\begin{figure}[t]
\centering
\includegraphics[width=0.4\textwidth]{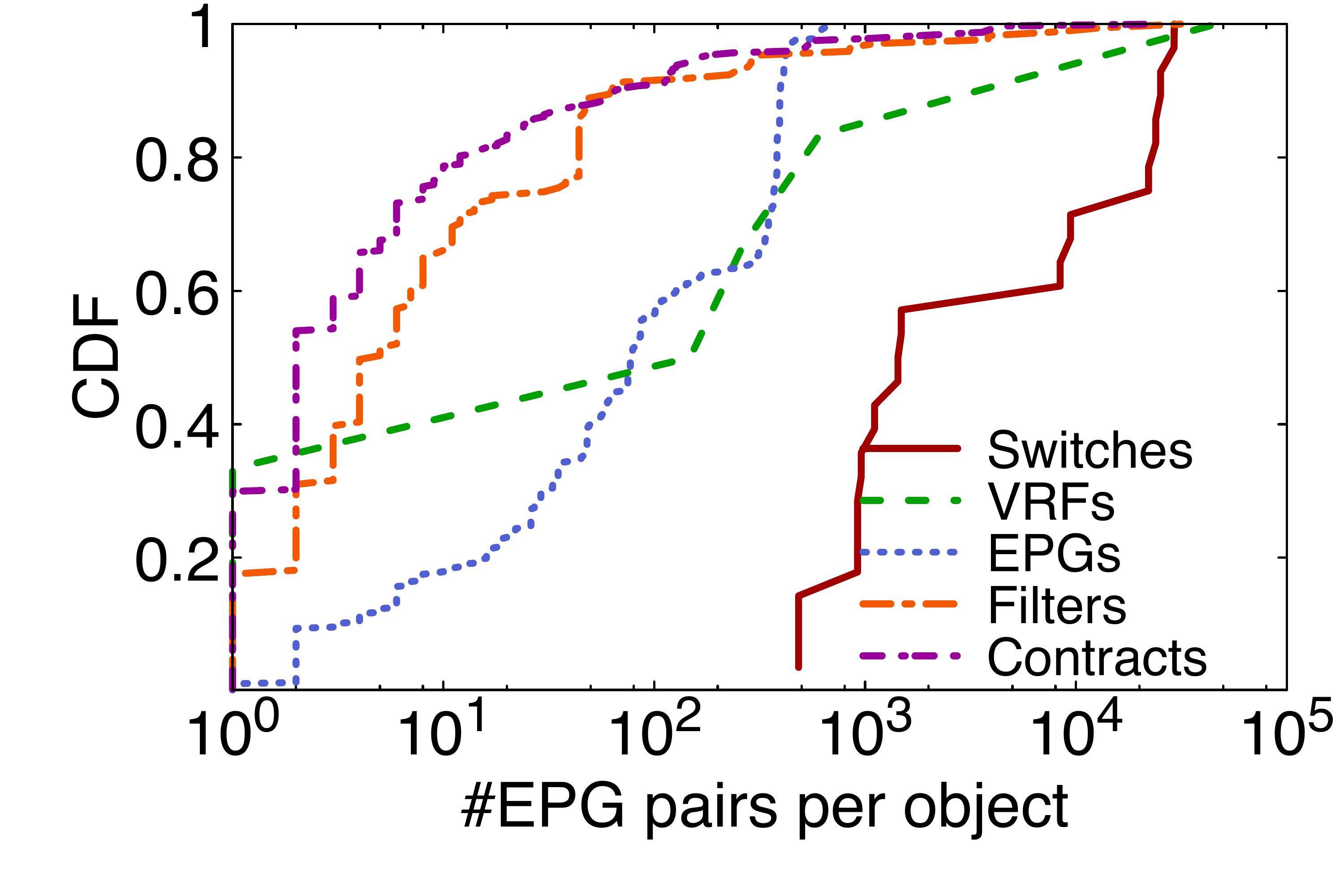}
\caption{Number of EPG pairs per object.}
\label{fig:cardinality}
\end{figure}

\subsection{A case study in a production cluster}
\label{s:cardinality}

A key aspect of shared risks is that they can create different degree of damages
to EPG pairs. If an incorrect VRF ID is distributed from the controller to
switch agents, all pairs of EPGs belonging to the VRF would be unable to
communicate. In contrast, if one filter is incorrectly deployed in one switch,
the impact would be limited to the endpoints in the EPG pairs that are directly
connected to the switch (and to other endpoints that might attempt to talk to
those endpoints).

In a network policy, a large number of EPG pairs may depend on a shared risk
(object) and/or a single EPG pair may rely on multiple shared risk
objects. These not only signify the criticality of a shared risk but also the
vulnerability of EPG pairs. More importantly, a dense correlation between shared
risks and EPG pairs makes it promising to apply risk modeling techniques to
fault localization of network policy deployment.

To understand the degree of sharing between EPG pairs and policy objects, we
analyze policy configurations from a real production cluster that comprises
about 30 Cisco's nexus 9000 series switches, one APIC, and hundreds of
servers. \fref{fig:cardinality} shows the cumulative distribution function on
the number of EPG pairs sharing a policy object, from which we make the
following observations:
\vspace{0.05in}
\begin{itemize}
\setlength\itemsep{0.5em}
\item \emph{A failure in deploying VRF would lead to a breakdown of a number of
    EPG pairs.} A majority of VRF objects has more than 100 EPG pairs. 10\% VRFs
  are shared by over 1,000 EPG pairs and 2-3\% VRFs by over 10,000 EPG pairs.

\item \emph{EPGs are configured to talk to many EPGs.} About 50\% of EPGs belong
  to more than 100 EPG pairs, which implies that the failure of an EPG is
  communication outage with a significant number of EPGs. 

\item \emph{The failure of a physical object such as switch would create the
    biggest impact on EPG pairs.} About 80\% of switches maintain at least
  1,000s of EPG pairs.

\item \emph{Contract and filter are mostly shared by a small number of EPG
    pairs.}
   70\% of the filters and 80\% of the contracts are used by less than 10 EPG
  pairs.
 
\end{itemize}

From these observations, it is evident that failures in a shared risk affect a
great number of EPG pairs. Consider a problematic situation where admins see
numerous high alerts that indicate a communication problem between a number of
EPG pairs, all because of a few shared risk failures. In this case, it is
notoriously hard to inspect individual EPG pairs and find out the underlying
faulty risk objects. However, this high dependency also makes spatial
correlation hold promise in localizing problematic shared risks among a huge
number of shared risks in large-scale networks.

\begin{figure}[t]
\centering
\subfigure[Switch risk model for switch $S_2$. When the 1st rule is missing from
  the TCAM in $S_2$ in \fref{fig:swrules}, the edges associated with the Web-App
  EPG pair are marked as fail (details in \secref{s:augment}).] {
\includegraphics[width=0.3\textwidth]{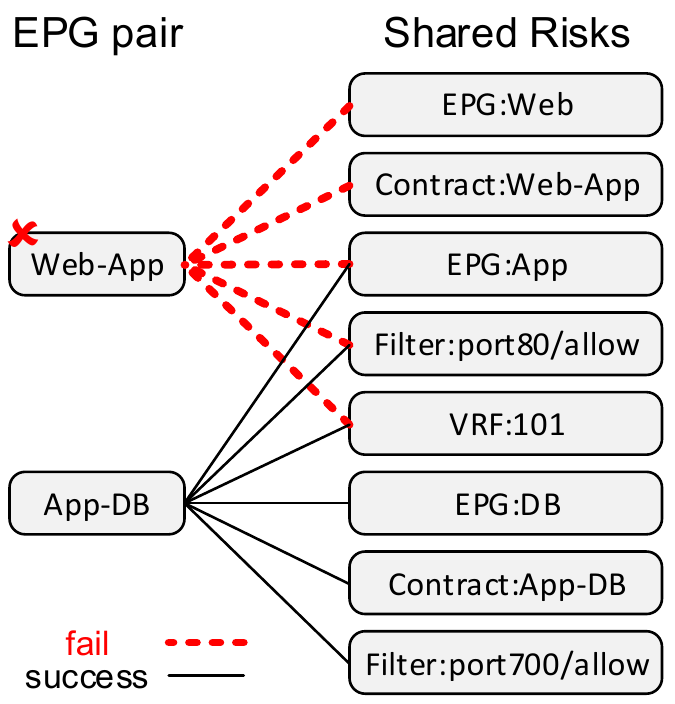} 
\label{fig:swrisk}}
\subfigure[Controller risk model for tenant network policy shown in
\fref{fig:bgexp}(b). Here, only edges associated with the $S_2$-Web-App are
marked as fail, because a rule (1st rule in \fref{fig:swrules}) is missing only
in $S_2$, but the corresponding rule is present in other switches $S_1$ and
$S_3$.]{
\includegraphics[width=0.3\textwidth]{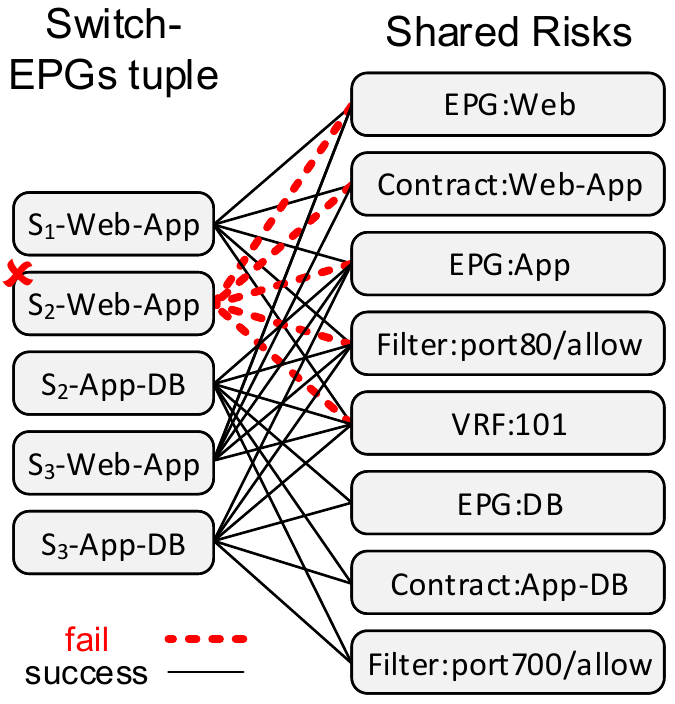}
\label{fig:ctrlrisk}}
\caption{Risk models for policy fault localization.}
\label{fig:risk}
\end{figure}

\subsection{Risk models}
\label{sec:risk_models}

We adopt a bipartite graph model that has been actively used to model risks in
the traditional IP network~\cite{score}.
A bipartite graph demonstrates associations between policy objects and the
elements that would be affected by those objects. At one side of the graph are
policy objects (\eg, VRF, EPG, filter, etc.); and the affected elements (\eg,
EPG pairs) are located at the other side. An edge between a pair of nodes in the
two parties is created if an affected element relies on a policy object under
consideration.

In modeling risks for network policy, one design question is how to represent
risks in the 3-tier deployment hierarchy that involves controller, switch agent
and TCAM. During rule deployment, there are two major places that eventually cause the
failure of TCAM rule update---one from controller to switch agent and the other
from switch agent to TCAM. The former may cause global faults whereas the latter
does local faults. For instance, if the controller cannot reach out to a large
number of switches for some reason, the policy objects across those unreachable
switches are not updated. On the other hand, when one switch is unreachable, a
switch agent misbehaves, or TCAM has hardware glitches, the scope of risk model
should be restricted to a particular switch level. Thus, in order to capture
global- and local-level risks properly, we propose two risk models: (i) switch
and (ii) controller risk model.

\myparab{Switch risk model.} A switch risk model consists of shared risks (\ie,
policy objects) and the elements (\ie, EPG pair) that can be impacted by the
shared risks on a per-switch basis. The model is built from a network policy and
the physical locations of endpoints belonging to EPGs in the network.
\fref{fig:swrisk} shows an example of switch risk model for switch $S_2$ given
the local view on network policy in \fref{fig:bgexp}(c). The left-hand side in
the model shows all EPG pairs deployed in switch $S_2$. Each EPG pair has an
edge to those policy objects (on the right-hand side in the model) that it
relies on in order to allow traffic between endpoints in the EPG pair. For
instance, the Web-App EPG pair has outgoing edges to EPG:Web, EPG:App, VRF:101,
Filter:port80/Allow, and Contract:Web-App. An edge is flagged as either success
or fail, soon discussed in \secref{s:augment}.

\myparab{Controller risk model.} A controller risk model captures shared risks
and their relationships with vulnerable elements across all switches in the network. 
A controller risk model is constructed in a similar manner of a switch risk
model. In the controller risk model, a switch ID and an EPG pair form a triplet.
A triplet has edges to policy objects that the EPG pair relies on in that
specific switch. Since the same policy object can be present in more than one
switch, an EPG pair in multiple switches can have an edge to the object.
\fref{fig:ctrlrisk} shows the controller risk model for tenant network policy
presented in \fref{fig:bgexp}(b).

\subsection{Augmenting risk models}
\label{s:augment}

In a conventional risk model, when an element affected by shared risks
experiences a failure, it is referred to as an \emph{observation}. In case of
switch risk model, an EPG pair is an observation when endpoints in the EPG pair
are allowed to communicate but fail to do so.

In our work, an observation is made by collecting the TCAM rules (T-type rules)
deployed across all switches periodically and/or in an event-driven fashion, and
by conducting an equivalence check between logical TCAM rules (L-type rules)
converted from the network policy at the controller and the collected T-type
rules. For this, we use an in-house equivalence checker.
The equivalence check is to compare two reduced ordered binary decision diagrams
(ROBDDs); one from L-type rules, and the other from T-type rules. If both ROBDDs
are equivalent, there is no inconsistency between the desired state (\ie, the
network policy) and actual state (\ie, the collected TCAM rules). If not, the
tool generates a set of \emph{missing} TCAM rules that explains the difference
and that should have been deployed in the TCAM but absent from the TCAM).
Those missing rules allow to annotate edges in the risk models as failure,
thereby providing more details on potentially problematic shared risks. Note
that simply reinstalling those missing rules is a stopgap, not a fundamental
solution to address the real problem that creates state inconsistency.

Potentially, the L-T equivalence checker can produce a large number of missing
rules. As demonstrated by our study on dependencies between objects
(\secref{s:cardinality} and \fref{fig:cardinality}), one ill-presented object at
controller and/or switch agent can cause policy violations for over thousands of
EPG pairs and make thousands of rules missing from the network. Unfortunately,
it is expensive to do object-by-object checking present in the observed
violations. Thus, we treat all objects in the observed violations as a potential
culprit. We then mark (augment) the edges between the \emph{malfunctioning} EPG
pair (due to the missing rule) and its associated objects in the violation as
fail.

\fref{fig:swrisk} illustrates how the switch risk model is augmented with
suspect objects if the 1st rule is missing from the TCAM in $S_2$ in
\fref{fig:swrules}.
To pinpoint culprit object(s), one practical technique is to pick object(s) that
explains the observation best (\ie, the famous Occam's Razor principle); in this
example, EPG:Web and Contract:Web-App would explain the problem best as they are
solely used by the Web-App EPG pair. The lack of the augmented data would make
it hard to localize fault policy objects as it suggests that all objects appear
equally plausible. Note that the example is deliberately made simple to ease
discussion. In reality many edges between EPG pairs and shared risks can be
marked as fail (again, see the high degree of dependencies between objects from
\fref{fig:cardinality}).

\section{Fault Localization}
\label{sec:faultloc}

We now build a fault localization algorithm that exploits the risk models
discussed in \secref{s:risk}. We first present a general idea, explain why the
existing approach falls short in handling the problem at hand
and lastly describe our proposed algorithm.

\subsection{General idea}
\label{s:idea}

In the switch risk model, for instance, an EPG pair
is marked as fail, if it has at least one failed edge between the pair and a
policy object (see \fref{fig:swrisk}). Otherwise, the EPG pair is success. Each
EPG pair node marked as fail is an \textit{observation}. A set of observations
is called a \textit{failure signature}. Any policy object shared across multiple
EPG pairs becomes a shared risk.

If all edges to an object are marked as fail, it is highly likely that the
failure of deploying that object explains the observations present in the
failure signature, and such an object is added to a set called
\emph{hypothesis}. Recall in \fref{fig:swrisk} that the EPG:Web and
Contract:Web-App objects best explain the problem of Web-App EPG pair. On the
other hand, other objects such as VRF:101 and EPG:App are less likely to be the
culprit because they are also shared by App-DB EPG pair which has no problem. An
ideal algorithm should be able to pick all the responsible policy objects as a
hypothesis.

In many cases, localizing problematic objects is not as simple as shown in
\fref{fig:swrisk}. Multiple object failures can take place simultaneously. In
such a case, it is prohibitive to explore all combinations of multiple objects
that are likely to explain all of the observations in a failure
signature. Therefore, the key objective is to identify a minimal hypothesis (in
other words, a minimum number of failed objects) that explains most of the
observations in the failure signature. An obvious algorithmic approach would be
finding a minimal set of policy objects that covers risk models presented as a
bipartite graph. This general set cover problem is known to be
NP-complete~\cite{min-set-cover}.

\subsection{Existing algorithm: \score}
\label{s:score}

We first take into account a greedy approximation algorithm used by
\score~\cite{score} system that attempts to solve the min set coverage problem
and that offers $O(\log{}n)$-approximation to the optimal
solution~\cite{score}, where $n$ is the number of affected elements (\eg,
EPG pairs in our problem).
We first explain the \score algorithm and further discuss its limitation.

\myparab{Algorithm.} The greedy algorithm in the \score system picks policy
objects to maximize two utility values---\emph{(i) hit ratio} and \emph{(ii)
  coverage ratio}---computed for each shared risk. We first introduce a few
concepts in order to define them precisely under our switch risk model. The same
logic can be applied to the controller risk model.

Let $G_i$ be a set of EPG pairs that depend on a shared risk $i$, $O_i$ be a
subset of $G_i$ in which EPG pairs are marked as fail (\textit{observations})
due to failed edges between the EPG pairs and the shared risk $i$, and $F$ be
the failure signature, a set of all observations, \ie, $F = \bigcup O_i$ for all
$i$. For shared risk $i$, a hit ratio, $h_i$ is then defined as:
\[ h_i = |G_i \cap O_i|/|G_i| = |O_i|/|G_i| \]
In other words, a hit ratio is a fraction of EPG pairs that are observations out
of all EPG pairs that depend on a shared risk.  A hit ratio is 1 when all EPG
pairs that depend on a shared risk are marked as fail. And a coverage ratio,
$c_i$ is defined as:
\[ c_i = |G_i \cap O_i|/|F| = |O_i|/|F| \]
A coverage ratio denotes a fraction of failed EPG pairs associated with a shared
risk from the failure signature.

The algorithm chooses shared risks whose hit ratio is above some fixed threshold
value. Next, given the set of selected shared risks, the algorithm outputs those
shared risks that have the highest coverage ratio values and that maximize the
number of explained observations.

\begin{algorithm}[t]
\caption{\system (F, R, C)}
\label{algo:localize}
\begin{algorithmic}[1]
\State $\triangleright$ $F$: failure signature, $R$: risk model, $C$: change logs
\State $\triangleright$ $P$: unexplained set, $Q$: explained set, $H$: hypothesis
\State $P \gets \texttt{F}$; $Q \gets \emptyset$; $H \gets \emptyset$

\While{$P \neq \emptyset$} \label{alg:candi_start}
        \State $K \gets \emptyset$ \Comment{$K$: a set of shared risks}
	\For{$observation$ $o \in P$} \label{alg:obs_start}
		\State $objs \gets \texttt{getFailedObjects}(o, R)$
		\State $\texttt{updateHitCovRatio}(objs, R)$
                \State $K \gets K \bigcup objs$
	\EndFor \label{alg:obs_end}

	\State $faultySet \gets \texttt{pickCandidates}(K)$ \label{alg:pick}
	\If{$faultySet = \emptyset$} 
	\State \textbf{break}
	\EndIf
        \State $affected \gets \texttt{GetNodes} (faultySet, R)$ \label{alg:prune_start}
        \State $R \gets \texttt{Prune}(affected, R)$ \label{alg:prune_end}
        \State $P \gets P \setminus affected$; $Q \gets Q \bigcup affected$ \label{alg:update}
        \State $H \gets H \bigcup faultySet$
\EndWhile \label{alg:candi_end}
\If{$P \neq \emptyset$}
	\For{$observation$ $o \in P$} \label{alg:left_start}
        \State $objs \gets \texttt{lookupChangeLog}(o, R, C)$
        \State $H \gets H \bigcup objs$
	\EndFor \label{alg:left_end}
\EndIf
\State \Return $H$
\end{algorithmic}
\end{algorithm}
\setlength{\textfloatsep}{0pt}

\myparab{Limitation.} The algorithm treats a shared risk with a small hit ratio
as noise and simply ignores it. However, in our network policy fault
localization problem, we observe that while some policy objects such as filter
have a small hit ratio ($\approx$ 0.01), they are indeed responsible for the
outage of some EPG pairs. The algorithm excludes those objects, which results in
a huge accuracy loss (results in \secref{s:results}).

It turns out that not all EPG pairs that depend on the object are present in the
failure signature. For instance, suppose that 100 EPG pairs depend on a filter,
which needs 100 TCAM rules. In this case, if one TCAM rule is missing, a hit
ratio of the filter is 0.01. This can happen if installing rules for those EPG
pairs is conducted with a time gap. For instance, 99 EPG pairs are configured
first, and the 100th EPG pair is a newly-added service, hence configured later.

To make it worse, in reality the hit ratio can vary significantly too. In the
previous example, if 95 TCAM rules are impacted, the hit ratio is 0.95. The wide
variation of hit ratio values can occur due to (1) switch TCAM overflow; (2)
TCAM corruption~\cite{everflow} that causes bit errors on a specific field in a
TCAM rule or across TCAM rules; and (3) software bugs~\cite{bugs} that modify
object's value wrong at controller or switch agent. While the \score algorithm
allows change of a threshold value to handle noisy input data, such a static
mechanism helps little in solving the problem at hand, confirmed by our
evaluation results in \secref{sec:eval}.

\begin{algorithm}[t]
\caption{pickCandidates(riskVector)}
\label{algo:hitCoverage}
\begin{algorithmic}[1]
\State $hitSet \gets \emptyset$
\State $maxCovSet \gets \emptyset$
\For{$risk$ $r \in riskVector$} \label{alg:hitratio_start}
	\If{$\texttt{hitRatio}(r) = 1$} 
		\State $hitSet \gets hitSet \bigcup \{r\}$
	\EndIf
\EndFor \label{alg:hitratio_end}
\State $maxCovSet \gets \texttt{getMaxCovSet}(hitSet)$ \label{alg:maxcov}
\State \Return $maxCovSet$
\end{algorithmic}
\end{algorithm}
\setlength{\textfloatsep}{5pt}
\setlength{\textfloatsep}{\textfloatsepsave}

\subsection{Proposed algoirthm: \algo}
\label{s:scoutalgo}

We propose \algo algorithm that actively takes into account policy objects
whose hit ratio percentage is less than 100\% and thus overcomes the limitation
of the \score algorithm. Basically, our algorithm also greedily picks the faulty
objects and outputs hypothesis that has a minimal set of objects most likely
explains all the observations in a failure signature.

\aref{algo:localize} shows the core part of our fault localization algorithm.
The algorithm takes failure signature $F$ and risk model $R$ as input. $F$ has a
set of observations, \eg, EPG pairs marked as fail in the switch risk model.
For each observation in $F$, the algorithm obtains a list of policy objects with
fail edges to that observation and computes the utility values (\ie, hit and
coverage ratios) for all those objects
(lines~\ref{alg:obs_start}-\ref{alg:obs_end}). Then, based on the utility
values of shared risks in the model, the algorithm picks a subset of the shared
risks and treats them as faulty (line~\ref{alg:pick} and
\aref{algo:hitCoverage}). In \aref{algo:hitCoverage}, if the hit ratio of a
shared risk is 1, the risk is included in a candidate risk set
(lines~\ref{alg:hitratio_start}-\ref{alg:hitratio_end}); and then from the set,
the shared risks that have the highest coverage ratio values are finally chosen;
\ie, a set of shared risks that covers a maximum number of unexplained
observations (line~\ref{alg:maxcov}).

If $faultySet$ is not empty, all EPG pairs that have an edge to any shared risk
in the $faultySet$ are pruned from the model
(lines~\ref{alg:prune_start}-\ref{alg:prune_end}), and failed EPG pairs
(observations) are moved from unexplained to explained (line~\ref{alg:update}).
Finally, all the shared risks in the $faultySet$ are added to the hypothesis
set, $H$. This process repeats until either there are no more observations left
unexplained or when $faultySet$ is empty.

Some observations may remain unexplained because the shared risks associated
with those observations have a hit ratio less than 1 and thus are not selected
during the above candidate selection procedure. To handle the remaining
unexplained observations, the \algo algorithm searches logs about changes made
to objects (which are obtained from the controller), and selects the objects to
which some actions are recently applied
(lines~\ref{alg:left_start}-\ref{alg:left_end} in \aref{algo:localize}).
Despite its simplicity, this heuristic makes huge improvement in accuracy
(\secref{s:results}).

\begin{figure}[t]
\centering
\includegraphics[width=0.49\textwidth]{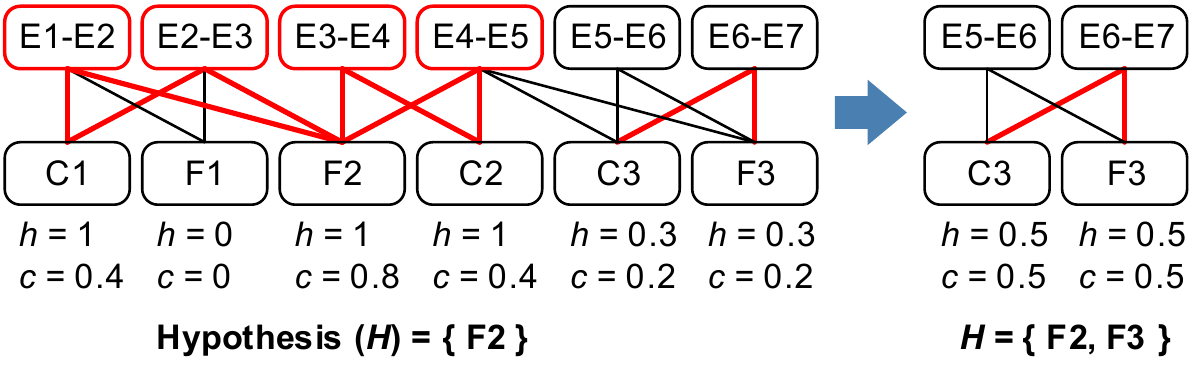}
\caption{An illustration of \algo algorithm using a switch risk model. Edges
and nodes in red color are fail and those in black are success. Note that $h$
refers to hit ratio and $c$ to coverage ratio.}
\label{fig:fault_loc_ex}
\end{figure}

\myparab{Example.} \fref{fig:fault_loc_ex} shows an example of how the \algo
algorithm works. The lines~\ref{alg:candi_start}-\ref{alg:candi_end} in
\aref{algo:localize} cover the following: (i) filter F2 is identified as a
candidate because it has the highest coverage ratio among the shared risks with
a hit ratio of 1; (ii) all the EPG pairs that depend on F2 are pruned from the
model; (iii) and F2 is added to hypothesis. The
lines~\ref{alg:left_start}-\ref{alg:left_end} ensure that the algorithm adds
filter F3 (assuming F3 is lately modified) to the hypothesis since there are no
shared risks with a hit ratio of 1.

\begin{figure*}[t]
\centering
\includegraphics[width=.98\textwidth]{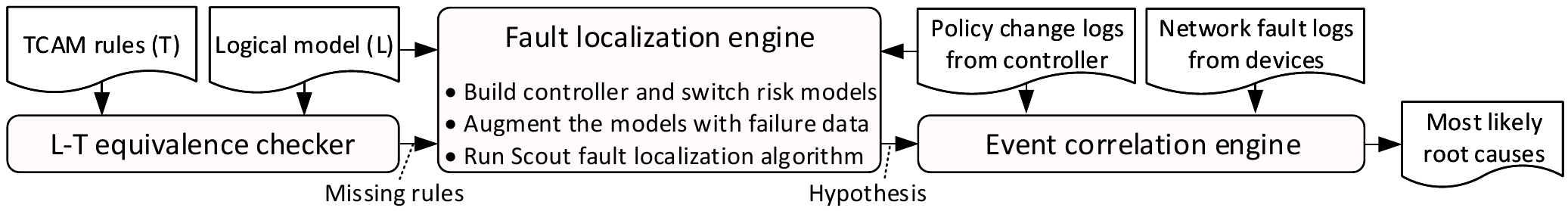}
\caption{Overview of \system system.}
\label{fig:system}
\end{figure*}

\section{\system System} 
\label{sec:arch}

We present \system system that can conduct an end-to-end analysis from fault
localization of policy objects to physical-level root cause diagnosis. The
system mainly consists of (i) fault localization engine and (ii) event
correlation engine. The former runs the proposed algorithm in
\secref{s:scoutalgo} and produces policy objects (\ie, hypothesis) that are
likely to be responsible for policy violation of EPG pairs. The latter
correlates the hypothesis and two system-level logs from the controller and
network devices, and produces the most-likely root causes at physical level that
caused object failures.
Our prototype is written in about 1,000 lines of Python code. We collect the
logical network policy model and its change logs from Cisco's application
centric controller, and switch TCAM rules and the device fault logs from Nexus
9000 series switches.
\fref{fig:system} illustrates the overall architecture of \system system.

\subsection{Physical-level root cause diagnosis}
\label{s:rca}

Knowing root causes at a physical level such as control channel disruption, TCAM
overflow, bugs, system crashes, etc. is as equally important as fixing failed
objects in the network policy. In general, when a trouble ticket is raised, the current practice is to 
narrow down possible root causes by analyzing system logs such as fault logs
from network devices. However, a majority in a myriad of log data is often
irrelevant to the caused failure. Filtering out such noises can be done to some
extent by correlating the logs with the generation time of the trouble ticket,
but not effective enough to reduce search space. 

To quickly narrow down the root cause of a failure, we rely on the hypothesis
reported by the fault localization algorithm.  We inspect only a small set of
failures (based on timestamps in fault logs) that are active around the same
time when changes are made to the faulty policy objects in the hypothesis.

The event correlation engine shown in \fref{fig:system} is a systematic and
automated approach to the above approach. The engine correlates the fault logs
from network devices, the change logs from the network policy controller and the
hypothesis generated from the fault localization engine. It then infers the most
likely physical-level root causes through the correlation.

The engine works in three simple steps: (i) Using the hypothesis, it first
identifies a set of change logs that it has to examine; (ii) with the timestamps
of those change logs, it then narrows down the relevant faulty logs (those that
are logged before the policy changes and keep alive); and (iii) it finally
associates impacted policy objects with the fault(s) found in the relevant
fault logs and outputs them.

The engine is pre-configured with signatures for known faults (\eg, disconnected
switch, TCAM overflow), composed by network admins with their domain knowledge
and prior experience. When fault logs match a signature, faults are identified
and associated with the impacted policy objects.  Otherwise, the objects are
tagged with `unknown'. Note that signatures can be flexibly added to the engine,
and the system's ability would be naturally enhanced with more signatures.

\subsection{Example usecases}
\label{sec:usecases}

We explain three realistic use cases in a testbed and demonstrate the workflow
of our system and its efficacy on fault localization. For this purpose, we use
the network policy for the 3-tier web service shown in \fref{fig:bgexp}(a). We
mimic a dynamic change of the network policy by continously adding one new
filter after another to the Contract:App-DB object. This would eventually cause
TCAM overflow. As a second case, we make a switch not respond to the controller
in the middle of updates, by silently dropping packets to the switch.

\myparab{TCAM overflow.} Due to TCAM overflow, several filters were not deployed
at TCAM. The switch under test generated fault logs that indicate TCAM overflow
when its TCAM utilization was beyond a certain level. Our system first localized
the faulty filter objects with risk models, correlated them with the change logs
for `add filter' instruction, and subsequently the change logs with the fault
logs. Our system had the fault signature of TCAM overflow, so it was able to
match the fault logs with that signature and tag those failed filters
accordingly.

\myparab{Unresponsive switch.} In this use case, the switch under test became
unresponsive while the controller was sending the `add filter' instructions to
the switch. The equivalence checker reported that the rules associated with some
filters are missing. Then, the \system algorithm localized those filters as
faulty objects. Using filter creation times from the change logs and the fault
logs that indicate the switch was inactive (both maintained at the controller),
the correlation engine was able to detect that filters were created when the
switch was inactive.

\myparab{Too many missing rules.} As a variant of the above scenario, we pushed
a policy with a large number of policy objects onto the unresponsive switch. We
found out that more than 300K missing rules were reported by the equivalence
checker.
Without fault localization, it is extremely challenging for network admins to
correlate and identify the set of underlying objects that are fundamentally
responsible for the problem. \system narrowed it down and reported the
unresponsive switch as the root cause behind these huge number of rule misses.

\section{Evaluation} 
\label{sec:eval}

We evaluate \system in terms of (i) suspect set reduction, (ii) accuracy, and
(iii) scalability. We mean by suspect set reduction a ratio, $\gamma$ between
the size of hypothesis (a set of objects reported by \system) and the number of
all objects that failed EPG pairs rely on; the smaller the ratio is, the less
objects network admins should examine. As for accuracy, we use precision
($|G\cap H|/|H|$) and recall ($|G\cap H|/|G|$) where $H$ is hypothesis and $G$
is a set for ground truth. A higher precision means fewer false positives and a
higher recall means fewer false negatives. Finally, we evaluate scalability via
measuring running times across different network sizes.

\subsection{Evaluation environment}

\myparab{Setup.} We conduct our evaluation under two settings.

\emph{Simulation}: We build our simulation setup with network policies used in
our production cluster that comprises about 30 switches and 100s of servers. The
cluster dataset contains 6 VRFs, 615 EPGs, 386 contracts, and 160 filters.

\emph{Testbed}: We build a network policy that consists of 36 EPGs, 24
contracts, 9 filters, and 100 EPG pairs, based on the statistics of the number
of EPGs and their dependency on other policy objects obtained from the above
cluster dataset.

\myparab{Fault injection.} We define two types of faults that cause
inconsistency between network policy and switch TCAM rules.
\emph{(i) Full object fault} means that all TCAM rules associated with an object
are missing.
\emph{(ii) Partial object fault} is a fault that makes some of the EPG pairs
that depend on an object fail to communicate. That is, some TCAM rules
associated with the object are missing.
For both simulation and experiment, we randomly generate the two types of faults
with equal weight.

\begin{figure}[t]
\centering
\subfigure[Faults in testbed]{
\includegraphics[width=0.5\textwidth,trim=0 20 0 20]{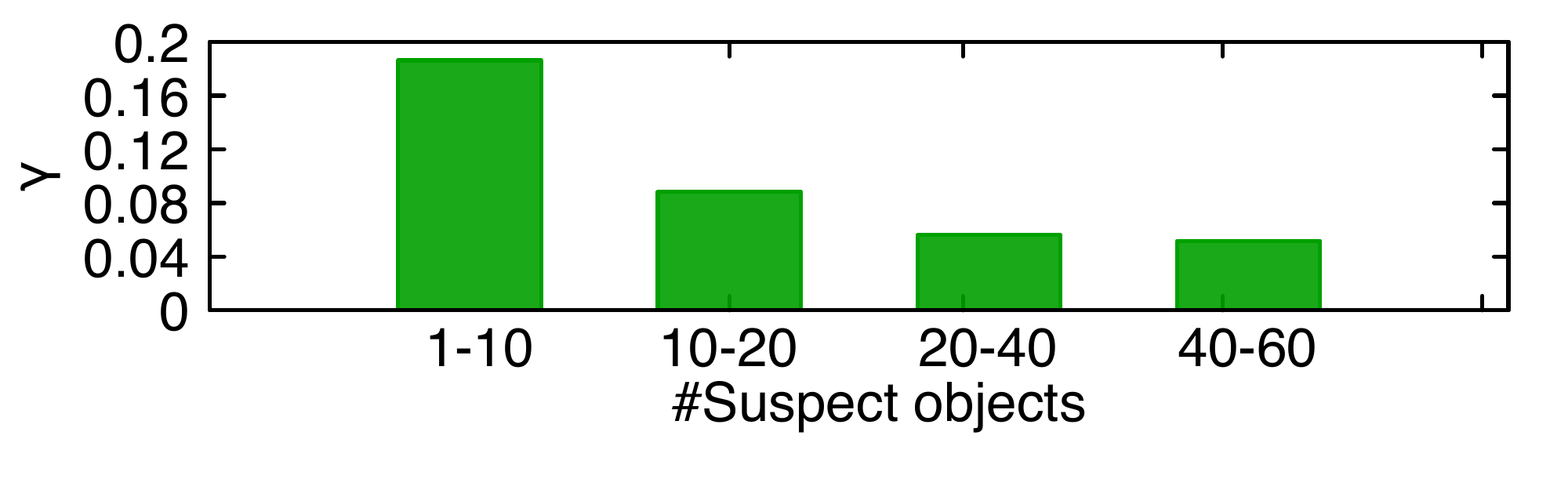}
\label{fig:faults_testbed}}
\subfigure[Simulated faults]{
\includegraphics[width=0.5\textwidth,trim=0 10 0 20]{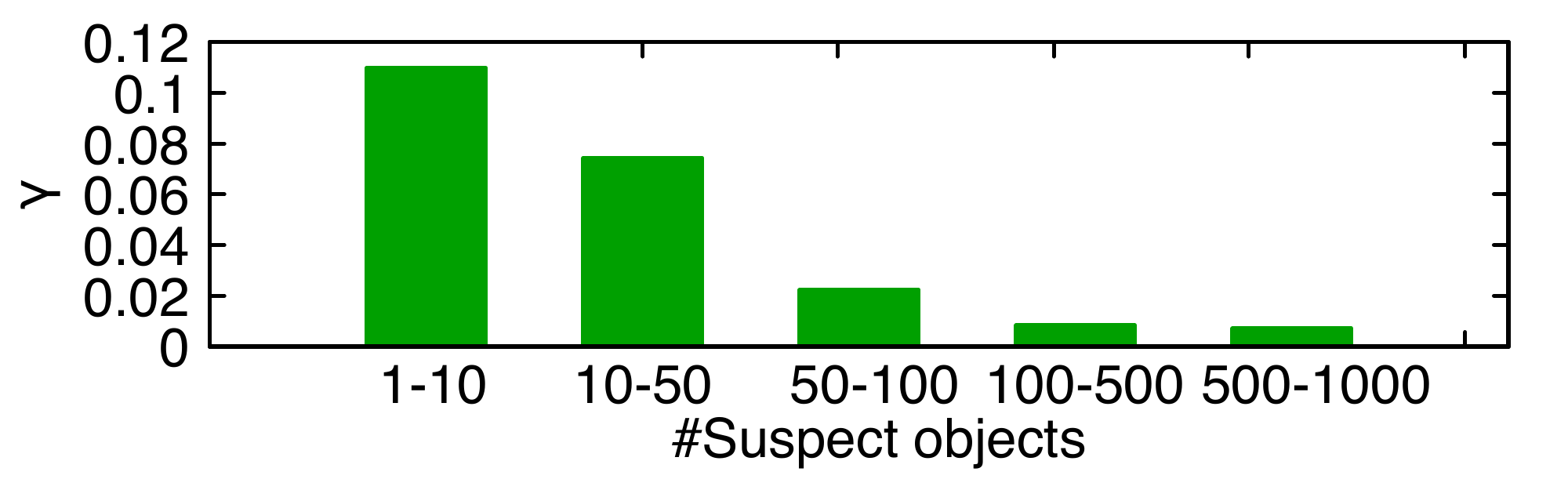}
\label{fig:faults_simulated}}
\caption{Suspect set reduction.}
\label{fig:suspect}
\end{figure}

\subsection{Results}
\label{s:results}
\myparab{Suspect set reduction.} We first compare the size of hypothesis with
the number of policy objects (a suspect set) that EPG pairs in failure depend
on.  We use the metric $\gamma$ defined earlier for this comparison.
\fref{fig:suspect} shows the suspect set reduction ratio in the simulation and
testbed. We generate 1,500 faults of object in the simulation and 200 faults of
object in the testbed; for each object fault, we compute the total number of
objects, that the EPG pairs impacted by the faulty object depend on. From the
figure, we see $\gamma$ is less than 0.08 in most cases. \system reports at
maximum 10 policy objects in the hypothesis whereas without fault localization
there are 
as many as a thousand policy objects to suspect. This smaller
$\gamma$ value means that network admins need to examine a relatively small
number of objects to fix inconsistencies between a network policy and deployed
TCAM rules. Therefore, \system can greatly help reduce repair time and necessary
human resources.

\begin{figure}[t]
\centering
\subfigure[Precision]{
\includegraphics[width=0.35\textwidth]{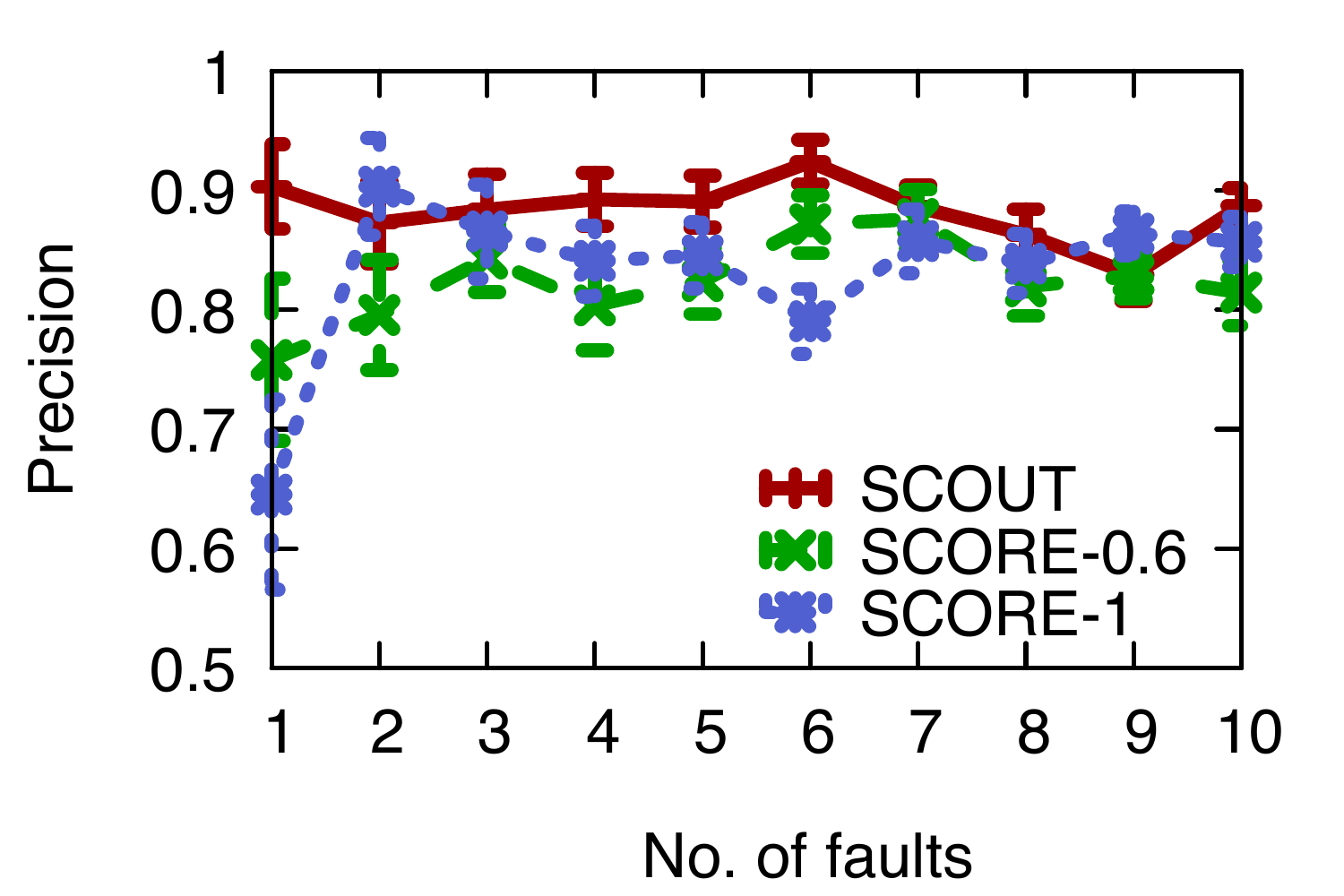}
\label{fig:switch_comb_precision}}
\subfigure[Recall]{
\includegraphics[width=0.35\textwidth]{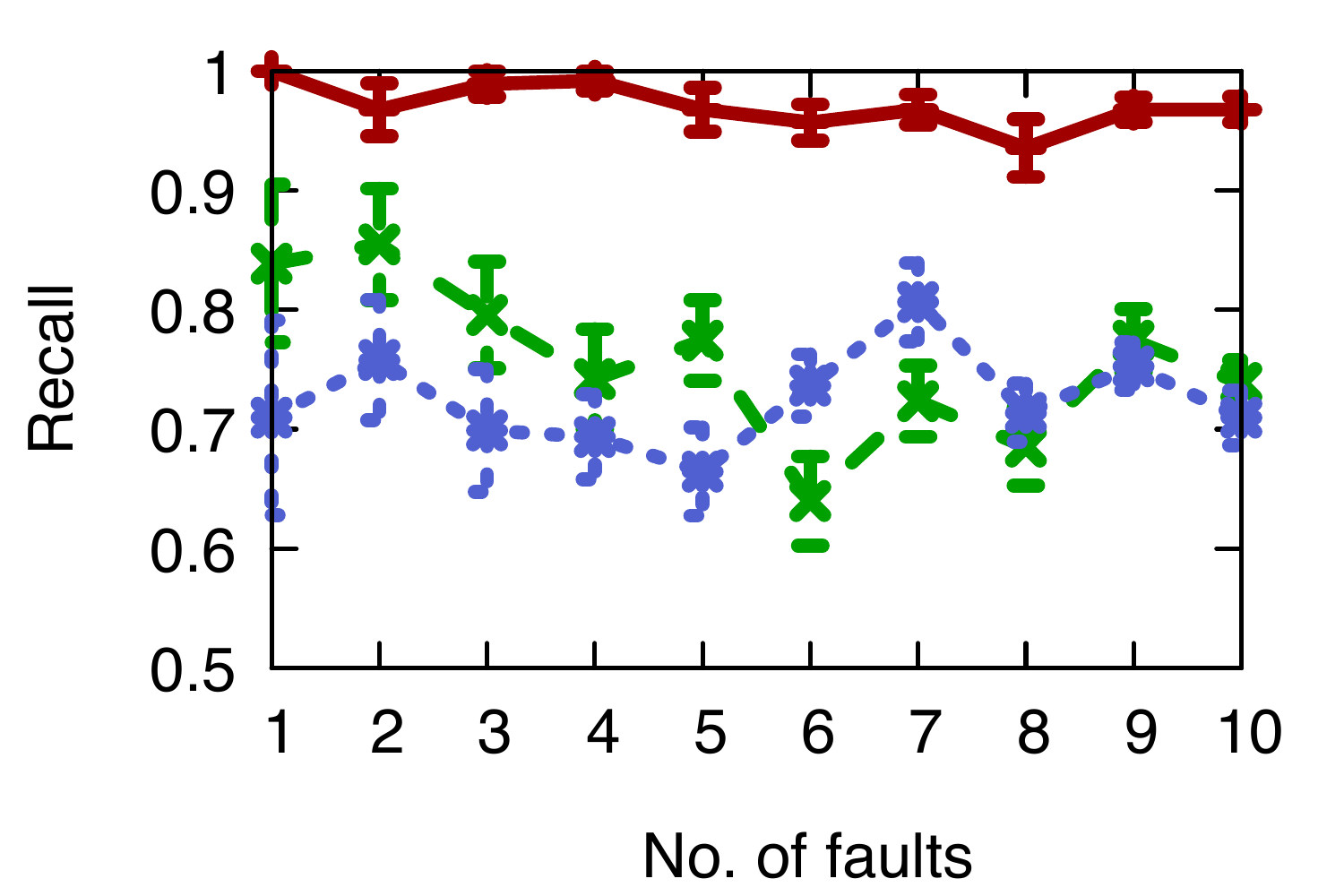}
\label{fig:switch_comb_recall}}
\caption{Fault localization performance on switch risk model. $X$ in \score-$X$
  is an error threshold set for hit ratio. The results are averaged over 30
  runs.}
\label{fig:switch_risk_model_comb}
\end{figure}

\begin{figure}[t]
\centering
\subfigure[Precision]{
\includegraphics[width=0.35\textwidth]{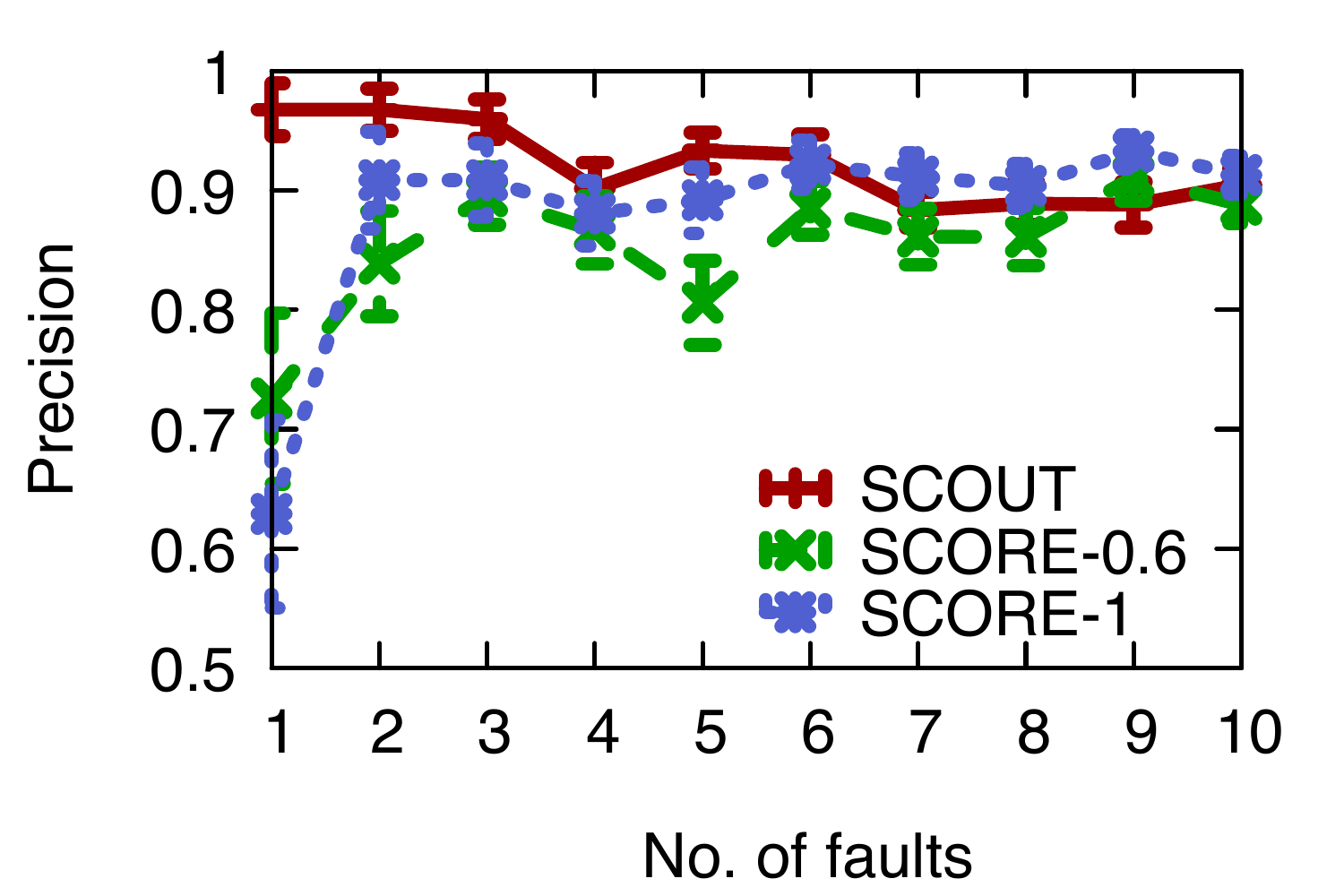}
\label{fig:controller_comb_precision}}
\subfigure[Recall]{
\includegraphics[width=0.35\textwidth]{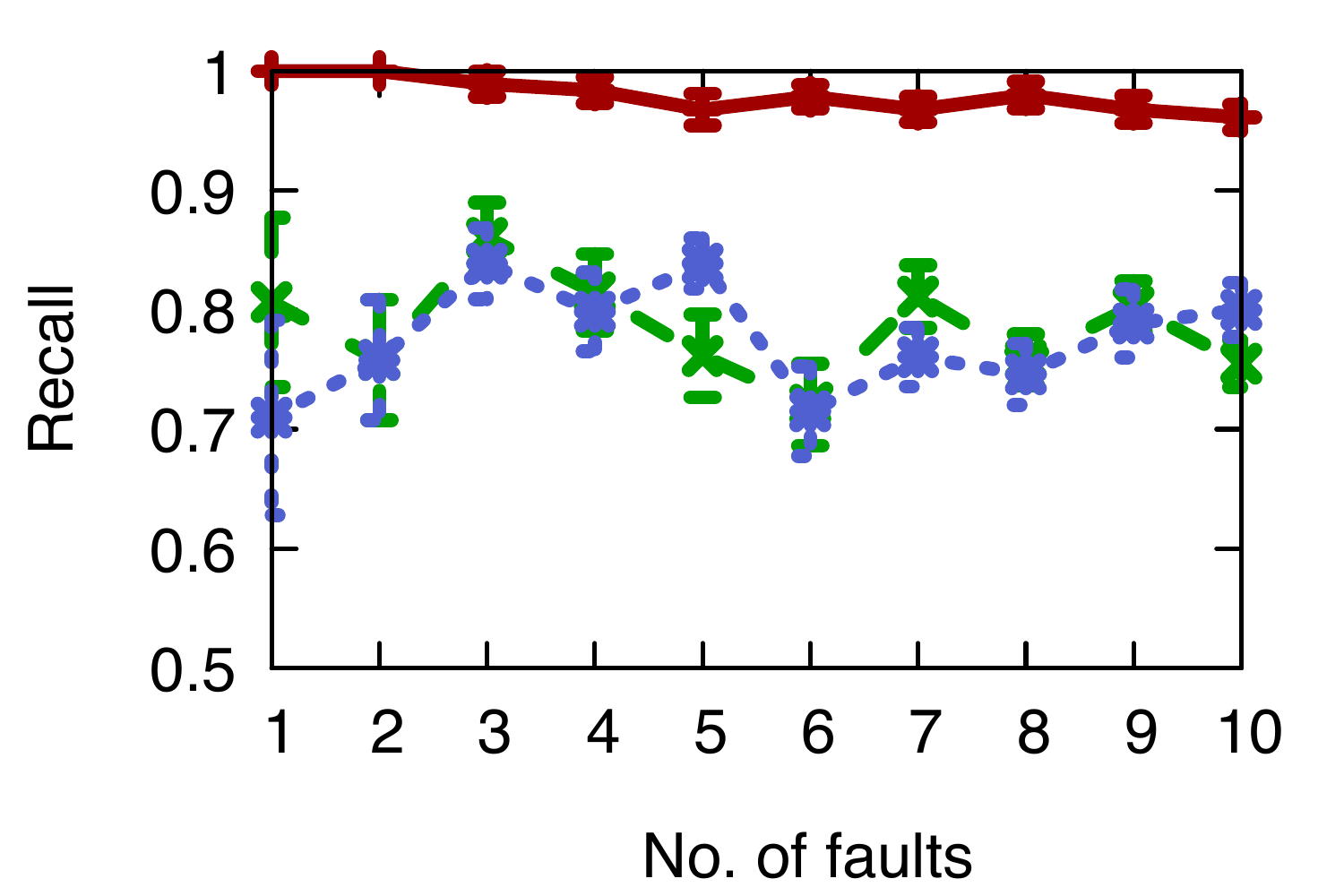}
\label{fig:controller_comb_recall}}
\caption{Fault localization performance on controller risk model with faulty
  policy objects across switches. 
  Each data point is an average over 30 runs.}
\label{fig:controller_risk_model_comb}
\end{figure}

\myparab{Accuracy.} While it is great that \system produces a handful of objects
that require investigation, a more important aspect is that the hypothesis
should contain more number of truly faulty objects and less number of non-faulty
objects. We study this using precision and recall. In addition, we compare
\system's accuracy with \score's. We use two different error threshold values
for \score to see if changing parameters would help improve its accuracy.

Figures~\ref{fig:switch_comb_precision} and \ref{fig:switch_comb_recall} show
recall and precision of fault localization with multiple number of simultaneous
faulty objects (x-axis) in the switch risk model. From the figures, we observe
\system's recall is 20-30\% better than \score's without any compromise on
precision. The error threshold values make little change in the performance of
\score.
Also, the high recall of \system suggests that \system can always find most
faulty objects. Moreover, high precision (close to 0.9) suggests fewer false
positives.
For instance, with 10 faulty objects in the network policy, \system reports on
average one additional object as faulty. In
Figures~\ref{fig:controller_comb_precision} and \ref{fig:controller_comb_recall}
we observe similar trends for the controller risk model.

\begin{figure}[t]
\centering
\subfigure[Precision]{
\includegraphics[width=0.35\textwidth]{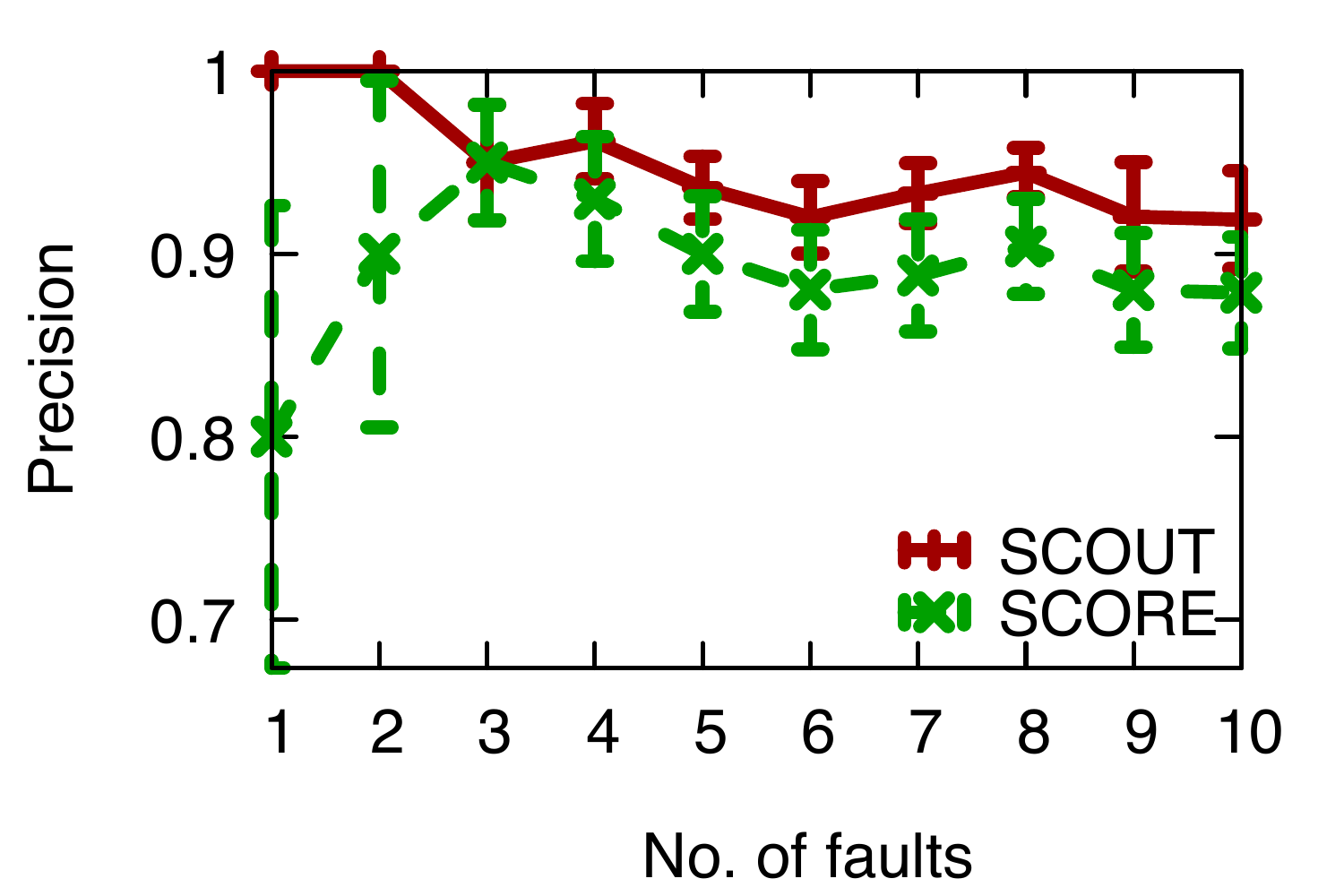}
\label{fig:switch_comb_precision_testbed}}
\subfigure[Recall]{
\includegraphics[width=0.35\textwidth]{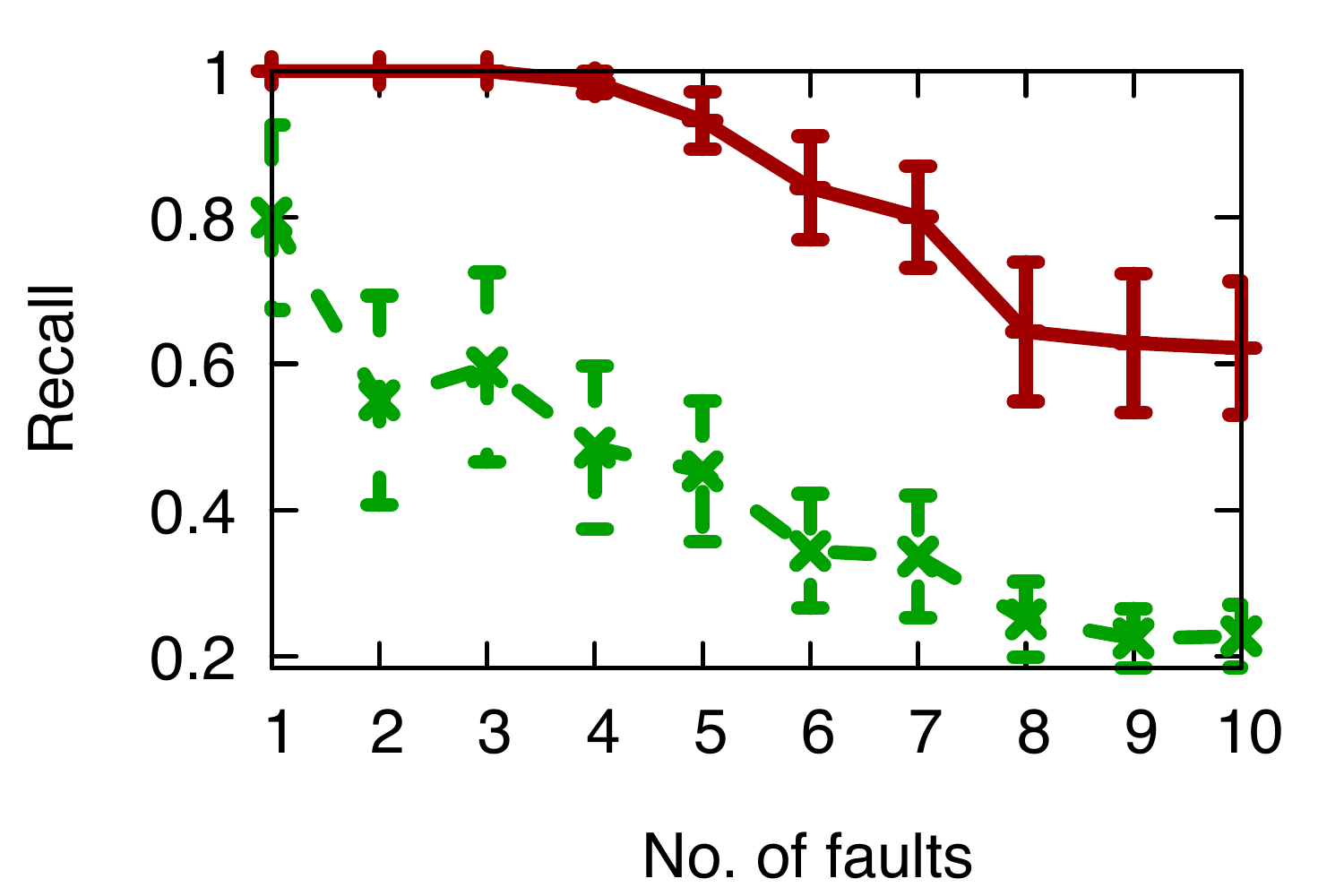}
\label{fig:switch_comb_recall_testbed}}
\caption{Fault localization performance when policy objects fail to be deployed
  in a switch. Each data point is an average over 10 runs.}
\label{fig:switch_risk_model_comb_testbed}
\end{figure}

Figures~\ref{fig:switch_comb_precision_testbed} and
\ref{fig:switch_comb_recall_testbed} compare the accuracy of \system and \score
with up to 10 simultaneous faults in the testbed. \score's error threshold is
set to 1. From the figures, we observe \system's recall is much better (20-50\%)
than \score's while its precision is comparable to \score's. \system detects all
faulty objects when there are less than four faults, \ie, with 100\% recall and
about 98\% precision. When there are five or more faults, \system's accuracy
(especially, recall) begins to decrease. The difference in accuracy between the
simulation and testbed setup is mainly because of a low degree of risk sharing
among EPG pairs in the testbed, when compared to the simulation dataset obtained
from the production cluster.

\begin{figure}[t]
 \centering
 \includegraphics[width=0.35\textwidth]{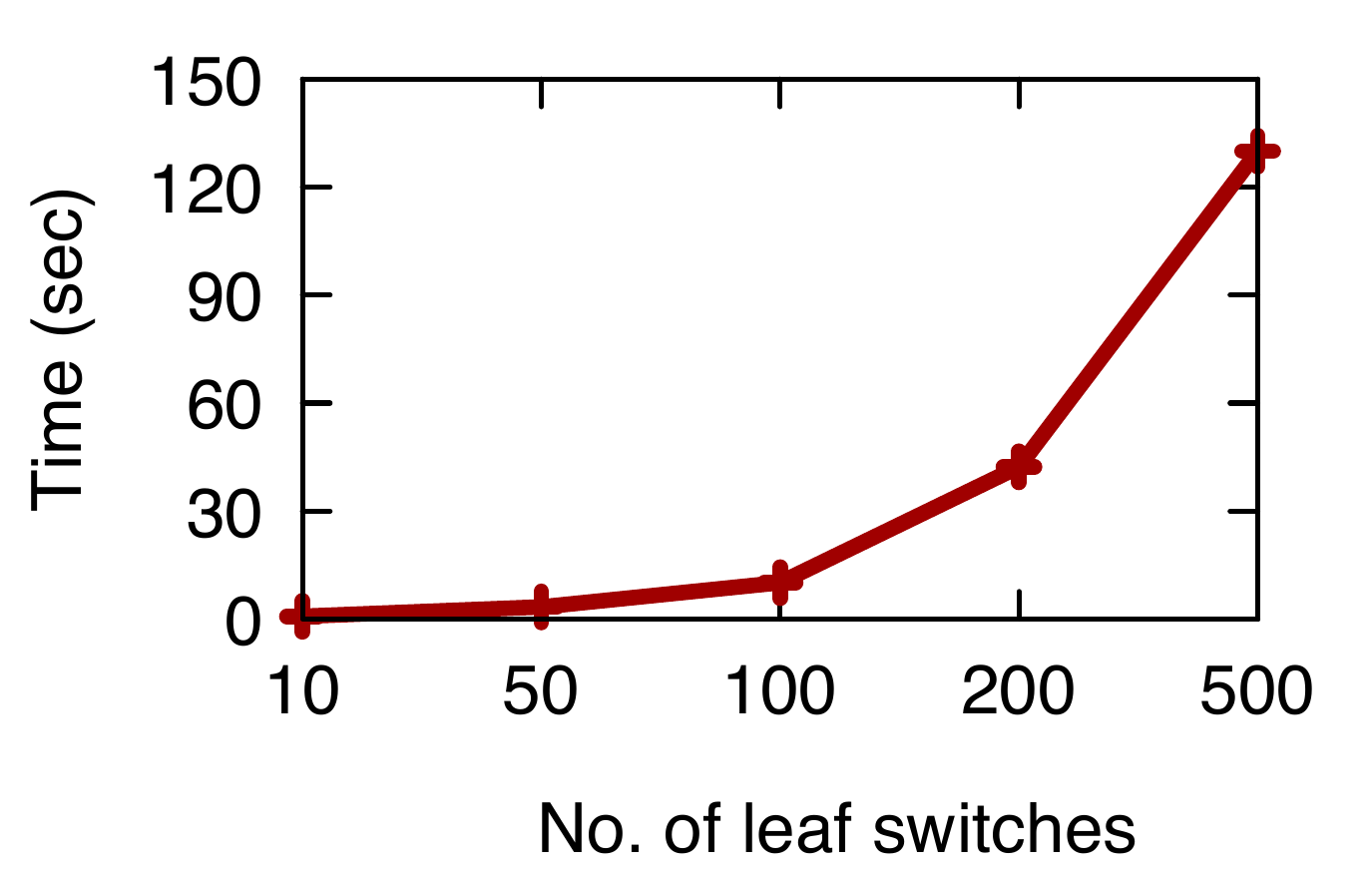}
 \caption{Localization execution time on controller risk model when number of faults are 10.}
 \label{fig:overhead}
 \end{figure}

\myparab{Scalability.} We measure running time of \system under a controller
risk model from the network policy deployed in 10 switches in the production
cluster. We scale the model up to 500 switches by adding new EPG and switch
pairs.
As shown in \fref{fig:overhead}, we observe that \system takes about 45 and 130 seconds with 200 and 500 switches respectively, on a machine with a 4-core 2.6 GHz CPU and 16GB memory.

\section{Related work} 
\label{sec:related_work}

A large body of research work has been conducted for network fault
localization~\cite{score, shrink, sherlock, gestalt, netdiagnoser, netmedic,
  blackholes, belief, faultloc}. Most of them focus on failures involving
physical components such as fibre-optic cable disruption, interface down, system
crash, etc. Our work focuses on fault localization of the network policy
configuration process rather than that of low-level physical components. Thus,
the context of our work is quite different from that of these prior works.

A number of recent approaches~\cite{pathquery, netsight, NetPlumber, veriflow,
  anteater, pathdump, HSA, everflow} enable data plane debugging.  Some of
them~\cite{pathquery, pathdump, everflow} collect debugging information by
installating low-level rules in the network switch either dynamically or
statically. PathQuery~\cite{pathquery} executes SQL-like queries on switches
while storing the query state in the packet header. PathDump~\cite{pathdump}
needs a small set of flow rules to embed path information into packet header.
Everflow~\cite{everflow} installs rules for selectively mirroring the
packets. In contrast, \system system does not install low-level rules to collect
debugging information. Instead, it relies on already available policy change
logs at the controller and failure logs at network switches.

Network provenance systems~\cite{netprov, diffprov} keep track of events
associated with packets and rules while \system only compares network policies
with actual rules deployed in the network.

Systems like Anteater~\cite{anteater}, Veriflow~\cite{veriflow}, and Header
Space Analysis~\cite{HSA} check for violation of invariants in network policies
on a data plane snapshot.  The snapshot includes ACLs and forwarding rules. In
particular, Anteater shares some similarity with the equivalence checker used by
\system in that both of them require access to TCAM rules. Unlike these systems,
\system focuses on localizing the part of the policy that is not deployed
correctly. Overall, \system compliments such approaches as it assists debugging
in management plane via fault localization.

Other works~\cite{pga, apic, frenetic, pyretic, merlin, participatory,
  onebigswitch, cocoon} focus on the automation of conflict-free, error-free composition
and deployment of network policies to reduce the likelihood of network problem
occurrences. While these frameworks are greatly useful in managing network
policies, it is hard to completely shield their management plane from failures,
which may cause the inconsistency between the policies and the actual network
state. \system can identify the impacted network policies. Thus, \system can be
useful in reinstating the network policies when these frameworks may not work
correctly.

\section{Conclusion}

Network policy abstraction enables flexible and intuitive policy management.
However it also makes network troubleshooting prohibitively hard when network
policies are not deployed as expected. In this paper we introduced and solved a
network policy fault localization problem where the goal is to identify faulty
policy objects that have low-level rules go missing from network devices and
thus are responsible for network outages and policy violations. We formulated
the problem with risks models and proposed a greedy algorithm that accurately
pinpoints faulty policy objects and built \system, an end-to-end system that
automatically pinpoints not only faulty policy objects but also physical-level
failures.

{
\balance
\bibliographystyle{IEEETran}
\bibliography{scout_tr}
}

\end{document}